\documentclass[aps,prl,amsmath,floats,floatfix, twocolumn,
superscriptaddress,nofootinbib,showpacs,reprint]{revtex4}

\usepackage{hyperref}

\usepackage{graphicx} 
\usepackage{xspace}
\usepackage[usenames,dvipsnames]{color}
\usepackage{amssymb}
\usepackage{longtable}

\newcommand{\NumPrecessing}{91\xspace}

\newcommand{\roughly}{\mathchar"5218\relax} 

\usepackage{ulem} 
\newcommand{\MakeTitle}{\maketitle}

\newcommand{\EarlyEnd}{}

\newcommand{\Caltech}{\affiliation{Theoretical Astrophysics 350-17,
    California Institute of Technology, Pasadena, CA 91125, USA}}
\newcommand{\Cornell}{\affiliation{Center for Radiophysics and Space
    Research, Cornell University, Ithaca, New York 14853, USA}}
\newcommand{\CITA}{\affiliation{Canadian Institute for Theoretical
    Astrophysics, 60 St.~George Street, University of Toronto,
    Toronto, ON M5S 3H8, Canada}} %
\newcommand{\CIFAR}{\affiliation{Canadian Institute for Advanced Research, 180 Dundas St.~West, Toronto, ON M5G 1Z8, Canada}} %
\newcommand{\DAA}{\affiliation{Department of Astronomy and Astrophysics, 50 St.\ George Street, University of Toronto, Toronto, ON M5S 3H4, Canada}}
\newcommand{\GWPAC}{\affiliation{Gravitational Wave Physics and
    Astronomy Center, California State University Fullerton,
    Fullerton, California 92834, USA}} %
\newcommand{\Oberlin}{\affiliation{Department of Physics and
    Astronomy, Oberlin College, Oberlin, Ohio 44074, USA}} %

\begin{document}

\title{
A catalog of 174 binary black-hole simulations for
gravitational-wave astronomy
}

\author{Abdul H.~Mrou\'{e}} \CITA 
\author{Mark A.~Scheel} \Caltech 
\author{B\'{e}la Szil\'{a}gyi} \Caltech 
\author{Harald P.~Pfeiffer} \CITA \CIFAR 
\author{Michael Boyle} \Cornell 
\author{Daniel A.~Hemberger} \Caltech \Cornell  
\author{Lawrence E.~Kidder} \Cornell 
\author{Geoffrey Lovelace} \GWPAC\Caltech 
\author{Serguei Ossokine} \CITA\DAA 
\author{Nicholas W.~Taylor} \Caltech 
\author{An\i l Zengino\u{g}lu} \Caltech 
\author{Luisa T. Buchman}\Caltech
\author{Tony Chu}\CITA 
\author{Evan Foley} \GWPAC 
\author{Matthew Giesler} \GWPAC 
\author{Robert Owen} \Oberlin
\author{Saul A. Teukolsky} \Cornell

\date{\today}

\begin{abstract}

  This paper presents a publicly available
  catalog of 174 numerical binary black-hole
  simulations following up to 35 orbits.  The
  catalog includes \NumPrecessing precessing
  binaries, mass ratios up to 8:1, orbital eccentricities from a few
  percent to $10^{-5}$, black-hole spins up 
    to 98\% of the theoretical maximum, and radiated energies up to 11.1\% 
  of the initial mass. 
  We establish remarkably good agreement with
  post-Newtonian precession of orbital and spin directions for two
  new precessing simulations, and we discuss other applications of this catalog. Formidable
  challenges remain: e.g., precession complicates the
  connection of numerical and approximate analytical waveforms, and
  vast regions of the parameter space remain unexplored.
\end{abstract}


\MakeTitle

\section{Introduction}


Gravitational waves (GW) from coalescing compact-object binaries --- 
neutron stars (NS) and stellar-mass black holes (BH) --- 
are primary targets for next-generation GW
detectors, such as Advanced LIGO, Virgo and
KAGRA~\cite{Harry:2010zz,aVIRGO,aVIRGO:2012,Somiya:2012}.  
Detecting GWs from compact-object binaries
requires
high-quality, accurate theoretical waveform models for
GW template banks.
Similarly,  
measuring source properties of detected 
signals (``parameter estimation'') relies on
theoretical waveform models 
used in Markov Chain Monte Carlo algorithms~\cite{Aasi:2013kqa}.

For widely separated binaries, post-Newtonian (PN)
calculations~\cite{Blanchet2006} provide accurate 
gravitational waveforms.
However,  numerical simulations of the full Einstein equations are 
needed 
during the late inspiral,
merger, and ringdown. 
Such simulations are particularly important for stellar-mass BH-BH and 
BH-NS systems: Late inspiral and merger occur near LIGO's most sensitive 
frequency range, and although BHs might have 
high spins~\cite{Gou:2011nq,McClintockEtAl:2006}, 
some of the spin contributions
to the PN waveforms are known only to lower
expansion order than the non-spinning terms 
(Ref.~\cite{2013arXiv1303.7412B} and references therein).


This paper focuses on binary black holes (BBH). 
BBH simulations 
became possible eight years
ago~\cite{Pretorius2005a}, with tremendous progress since
(e.g.,~\cite{Centrella:2010,Pfeiffer:2012pc}).  
For best utility to GW astronomy, such 
simulations 
must achieve 
(i) sufficient accuracy; (ii) a desired orbital
eccentricity, for instance the very small eccentricity expected for 
field binaries~\cite{PetersMathews1963,Peters1964}; (iii) sufficient
length (i.e., number of orbits) 
to connect reliably to PN waveforms; (iv) sufficiently dense
coverage of relevant regions of parameter space.  

Satisfying all conditions
(i) to (iv) is so difficult that simulations have barely
reached the minimal desired quality.
GW {\em detection} 
[condition (iii)]
requires 10 orbits for straightforward simulations (mass ratio
$q\lesssim 4$, dimensionless spins $\chi\equiv S/M^2\lesssim 0.7$
aligned with the orbital angular momentum)~\cite{OhmeEtAl:2011};
{\em parameter estimation} 
can benefit from well over
100 orbits~\cite{OhmeEtAl:2011,MacDonald:2012mp,Boyle:2011dy,Damour:2010}.
These numbers increase with more extreme mass ratio
and BH spins~\cite{Boyle:2011dy}, and analogous estimates have not even been 
performed for precessing binaries.  Yet few
published simulations follow even 10 orbits.

Additionally, the 7-dimensional
parameter space has barely been explored.
The world-wide
NINJA-2
collaboration~\cite{Ajith:2012az} yielded 
40 waveforms
with an
average length of $\roughly 9$ orbits, 
covering two
one-dimensional subspaces of the aligned spin region of 
the parameter 
space.  
The NRAR collaboration~\cite{Hinder:2013oqa} 
presented 25 waveforms of average length $\roughly 13$ orbits and
better accuracy than NINJA-2.
Pekowsky~et al.~\cite{Pekowsky:2013ska}
report about 80 simulations covering
approximately 7--12 orbits ($\roughly 25$ of them representing
precessing binaries) and $\roughly 170$
simulations lasting a few orbits each.


\begin{figure}[t]
\includegraphics[width=\columnwidth, bb=40 4 460 330, clip=true]{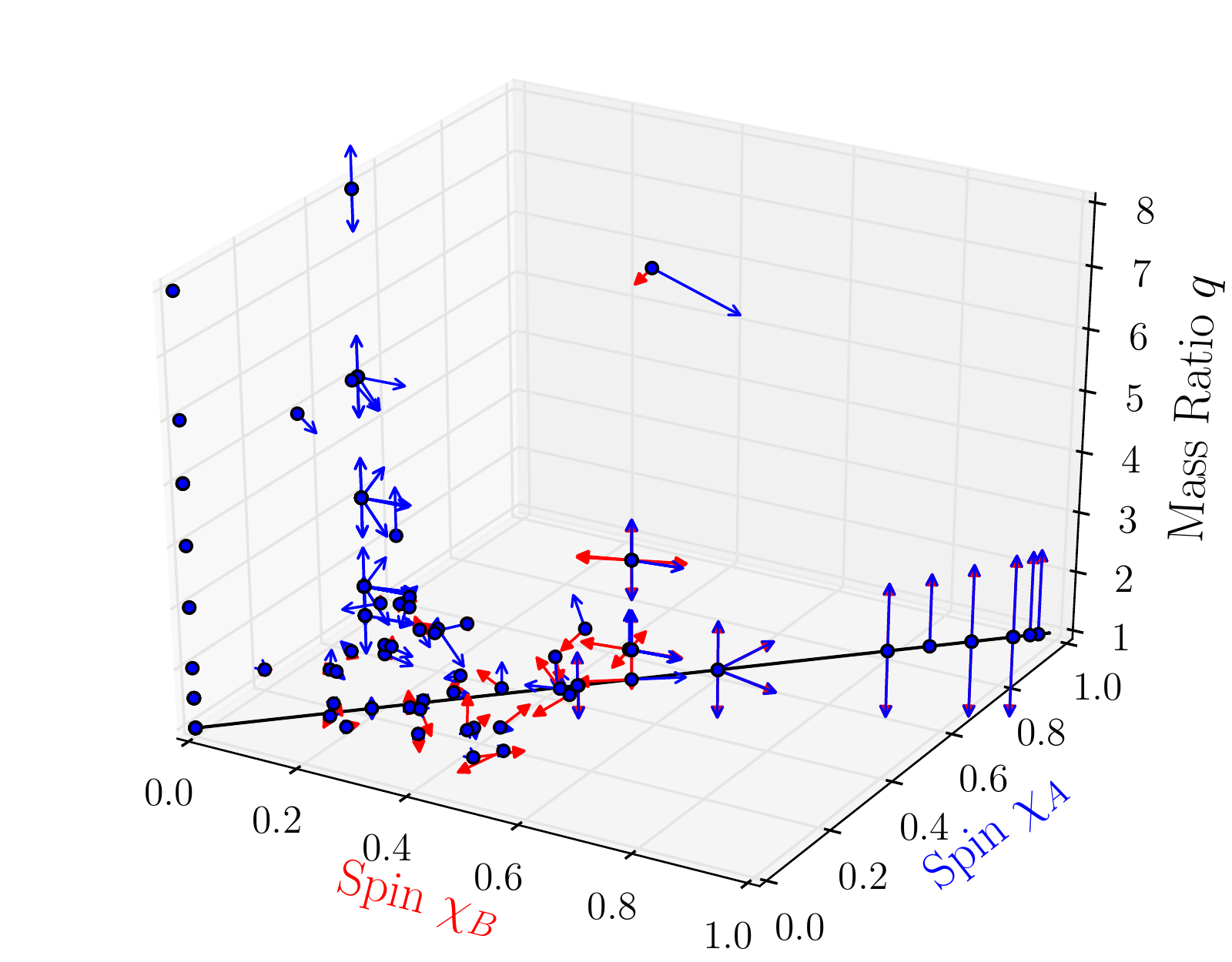}
\caption{The parameter space covered by our catalog. Each
point represents a simulation. The axes show the mass ratio 
and the spin magnitudes of
the larger (blue) and smaller (red) BHs. Arrows indicate the magnitude and 
direction of the spins. \label{fig:paramspace}}
\end{figure}

\begin{figure*}
\includegraphics[width=0.98\textwidth,trim=15 3 10 3,clip]{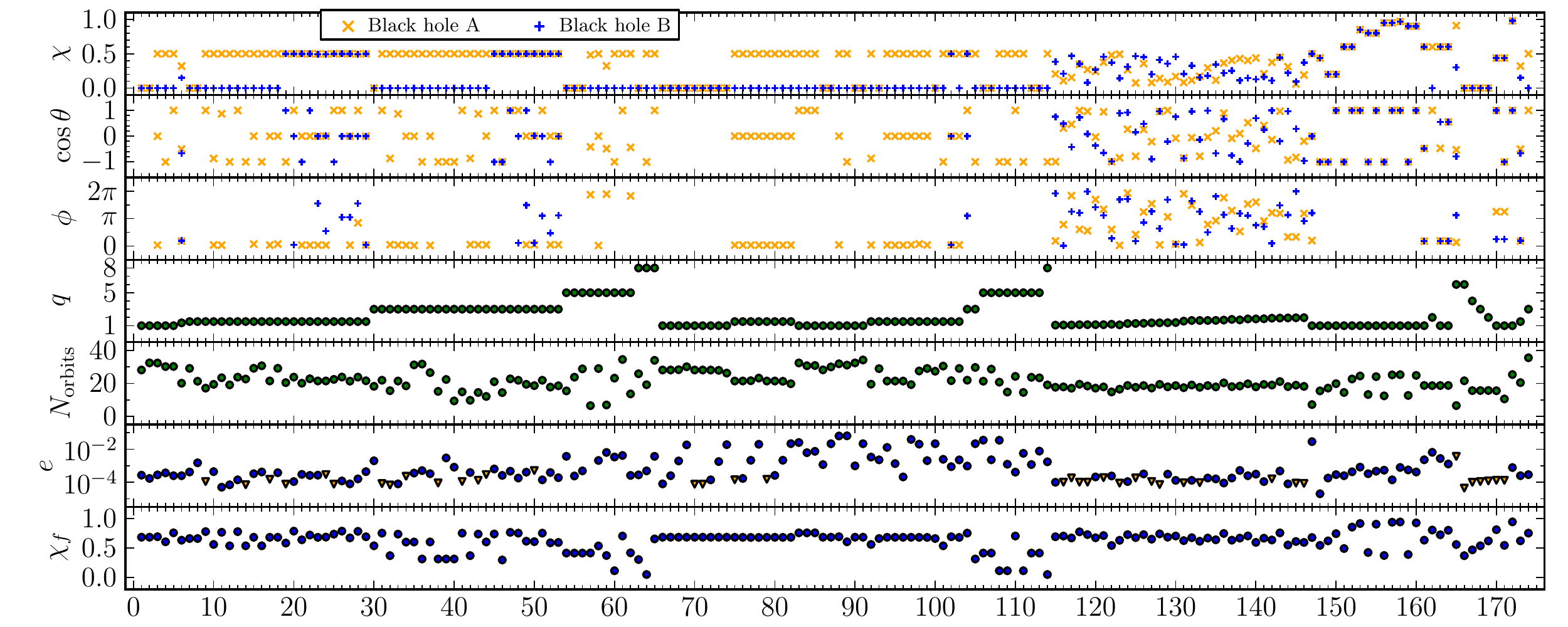}
  \caption{
    Properties of all 
    simulations in the catalog.  From
    top to bottom: 
    dimensionless initial spin
    magnitudes $\chi_{A,B}$; angles $\theta_{A,B}$
    between the initial spin vectors and the initial orbital angular momentum;
    angles $\phi_{A,B}$ between the line segment 
    connecting the centers of the black holes and the
    initial spin vectors projected into the
    initial orbital plane; mass ratio $q=M_A/M_B$; 
    number of orbits before merger;
    initial eccentricity $e$ (triangles indicate an upper bound on $e$); 
    final remnant spin $\chi_f$.
  }
\label{fig:listofruns}
\end{figure*}

\begin{figure*}[t]
\includegraphics[width=\textwidth,bb=5 5 964 533,clip=true]%
{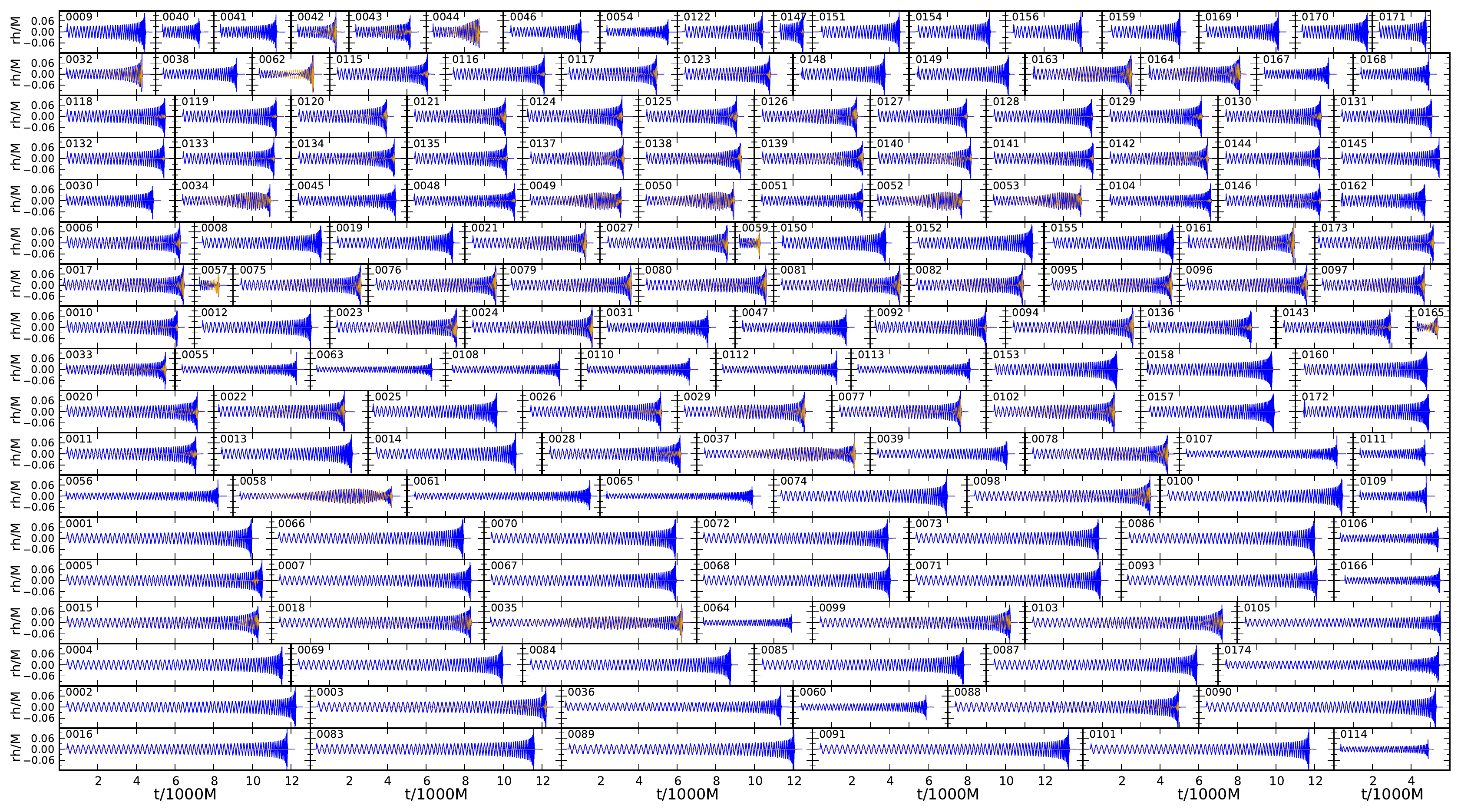}

\caption{Waveform polarizations $(r/M) h_+$ (blue) and $(r/M) h_\times$ (orange)
  in a sky direction parallel to the initial orbital plane of each
  simulation.  The unit of the time axis corresponds to $1000M=0.1s$ for binaries with total mass $M=20M_\odot$.
}
\label{fig:allwaveforms}
\end{figure*}

This paper represents a
major advance in waveform
length and parameter-space coverage.
We follow our earlier
approach~\cite{Buchman:2012dw,Lovelace:2011nu,Scheel2009,Boyle2007} of
computing BBH simulations 
that are longer than previous 
simulations, 
with higher accuracy and 
very low eccentricity.
Our catalog contains $174$ simulations (155
are new 
and previously unpublished).  
167 simulations cover more than
12 orbits (up to 35.5 orbits, Fig.~\ref{fig:listofruns}), and \NumPrecessing
represent precessing binaries.

This simulation catalog is
publicly available~\cite{SXSCatalog}  
and will have other applications besides its primary goal of facilitating
GW astronomy.

\section{Techniques}
The simulations are computed using the Spectral
Einstein Code ({\tt SpEC})~\cite{SpECwebsite}.  
Quasi-equilibrium~\cite{Caudill-etal:2006,Lovelace2008} initial
data are constructed~\cite{Pfeiffer2003} to solve the Einstein constraint 
equations~\cite{York1999} for binaries with low ($\roughly 10^{-4}$)
eccentricity~\cite{Pfeiffer-Brown-etal:2007,Buonanno:2010yk,Mroue:2012kv} 
and are evolved using a generalized
harmonic 
formulation~\cite{Friedrich1985, Garfinkle2002, Pretorius2005c,Lindblom2006}
of Einstein's equations and damped harmonic 
gauge~\cite{Lindblom2009c,Choptuik:2009ww,Szilagyi:2009qz}.
The adaptively-refined~\cite{Lovelace:2010ne} grid
extends from pure-outflow
excision boundaries conforming to
the shapes of the apparent
horizons~\cite{Scheel2009,Szilagyi:2009qz,Hemberger:2012jz,Ossokine:2013zga}
to an artificial outer boundary where constraint-preserving
boundary conditions~\cite{Lindblom2006, Rinne2006, Rinne2007} are imposed.
After merger, the grid has only a single excision 
boundary~\cite{Scheel2009,Hemberger:2012jz}.
Gravitational wave extraction and
extrapolation of waveforms to infinity are described
in~\cite{Pfeiffer-Brown-etal:2007, Boyle2007,
Rinne2008b, PanEtAl:2011, Buonanno:2009qa,Boyle-Mroue:2008,Boyle:2013a}.

No fine-tuning or trial-and-error is required for our current
evolution method, for regions of parameter space covered by this
catalog. Mesh refinement and grid transitions (such as 
re-gridding to a single excision boundary at merger)
are automatic.  This enables the completion of many simulations with
little cost in human time.
  
\begin{figure}
\includegraphics[width=0.97\columnwidth,trim=5 5 5 5]{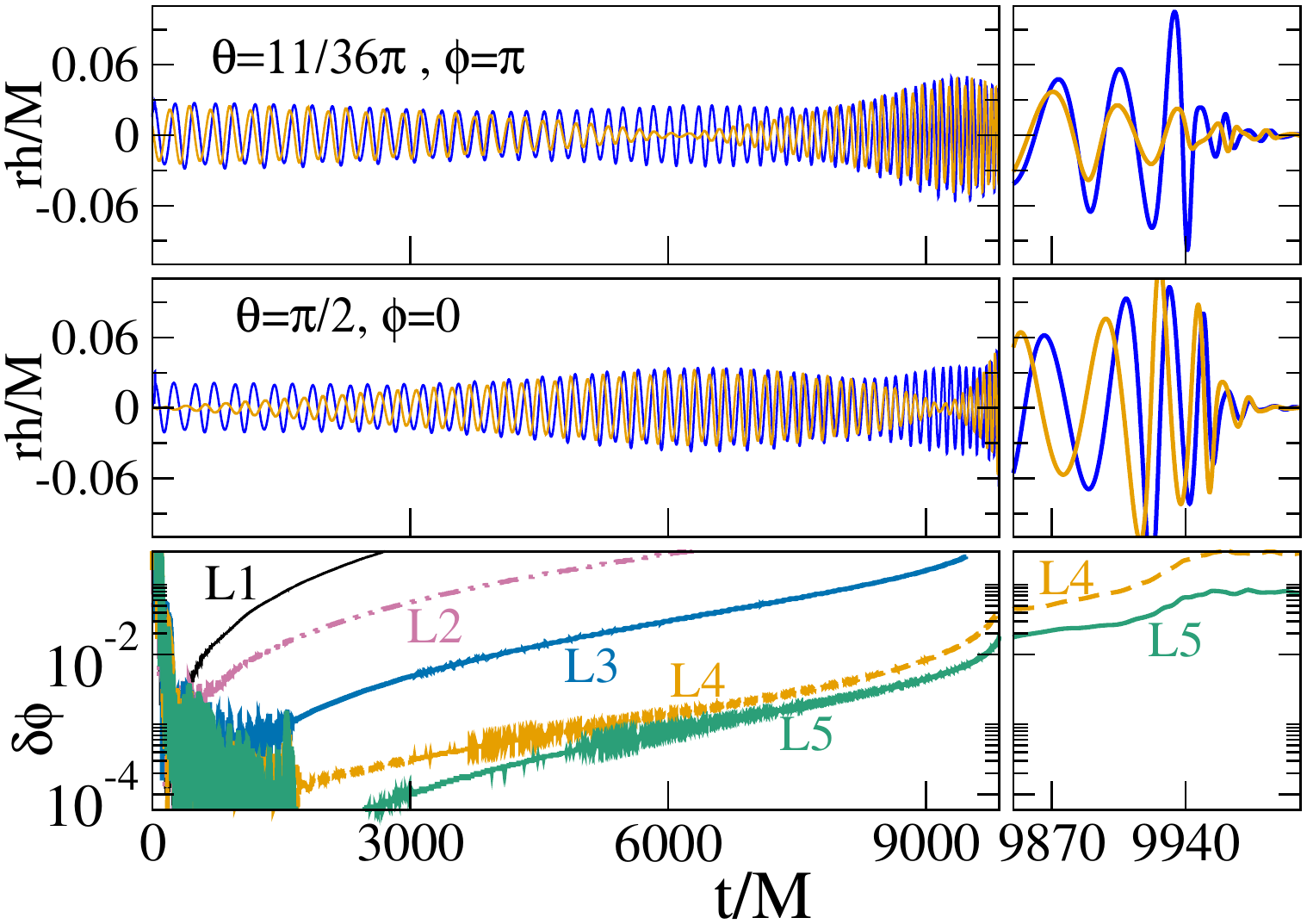}
\caption{\label{fig:convergence} {\bf Top:} waveforms $rh_+/M$ (blue) and
  $rh_\times/M$ (orange) for two detector 
  directions $(\theta,\phi)$ relative to the initial orbital 
  plane. {\bf Bottom:} 
cumulative phase differences between the $\ell\!=\!2,m\!=\!2$
  spherical-harmonic mode of the waveform of numerical
  resolutions L1 through L5
  compared to that of the highest resolution L6. 
}
\end{figure}

\section{Catalog}

Figure~\ref{fig:paramspace} represents the
  parameter space coverage.
  The catalog includes 
  (i) 62 new single-spin simulations with $\chi_A=0.5,
  \chi_B=0$ ranging from $q=1$ to $q=8$,  
  which extend 
  efforts
  by the NINJA-2 collaboration~\cite{Ajith:2012az} (which
  contains only three unequal mass,
  aligned spin simulations and no precessing simulations) and the
  NRAR collaboration~\cite{Hinder:2013oqa} (which contains
  spinning simulations only for
  $q\leq 3$);  (ii) 32 new simulations 
  with random mass ratios 
  $q\in[1,2]$ and
  random spins ($\chi_A,\chi_B\le 0.5$); (iii) 16 $q=1$ simulations with equal,
 aligned spins~\cite{Chu2009,Lovelace:2011nu,Hemberger:2013hsa}, 
 including one new simulation with the highest BH spin
 to date $(\chi_A\!=\!\chi_B\!=\!0.98$);
(iv) a high mass ratio,  
  high-spin ($q\!=\!6, \chi_A\!=\!0.9, \chi_B\!=\!0.3$), precessing simulation 
  intended
  to test numerical capabilities;   
  (v) 32 non-spinning simulations, including
  mass ratios $q\!=\!1.5, 5, 8$, not considered in~\cite{Buchman:2012dw}.
Also included are new simulations of different orbital eccentricities and durations,
which facilitate calculation of periastron advance~\cite{Mroue2010} and
efficient initial data generation~\cite{Mroue:2012kv}.

While we generally reduce orbital eccentricity~\cite{Buonanno:2010yk} to $e\lesssim 10^{-4}$,
 the catalog contains 43 new
  simulations with eccentricities from $e<10^{-4}$ up to
  $e=0.06$,  representing 
 the first systematic
 sampling of
 slightly eccentric BBH waveforms.

Figure~\ref{fig:allwaveforms} plots the gravitational-wave
 polarizations $h_+$ and $h_\times$ emitted into a
  certain sky direction, chosen so that $h_\times$ vanishes for non-precessing systems.  

Figure~\ref{fig:convergence} shows the waveforms for simulation $0035$ 
($q\!=\!3$, precessing, 31 orbits)
in~two sky directions, highlighting the waveforms' strong dependence 
on the BBH orientation relative to the line of
sight to Earth.  
Figure~\ref{fig:convergence} also presents a convergence test, 
showing differences in the waveforms computed using different
numerical resolutions.  
We reach {\em cumulative} phase errors similar to shorter {\tt SpEC} 
simulations~\cite{Boyle2007,Scheel2009,Buchman:2012dw},
which were instrumental for various GW data-analysis 
applications~\cite{PanEtAl:2011,Pan:2011gk,Ajith:2012az,Aylott:2009ya}.
The achieved error is about an
order of magnitude smaller than the NRAR project's target~\cite{Hinder:2013oqa} 
and is suitable for detection and parameter estimation.
This accuracy is representative for 
all simulations; a more detailed discussion will be presented 
in~\cite{Mroue:2013InPrep}.

\begin{figure}
\includegraphics[width=\columnwidth,bb=25 6 355 165,clip=true]
{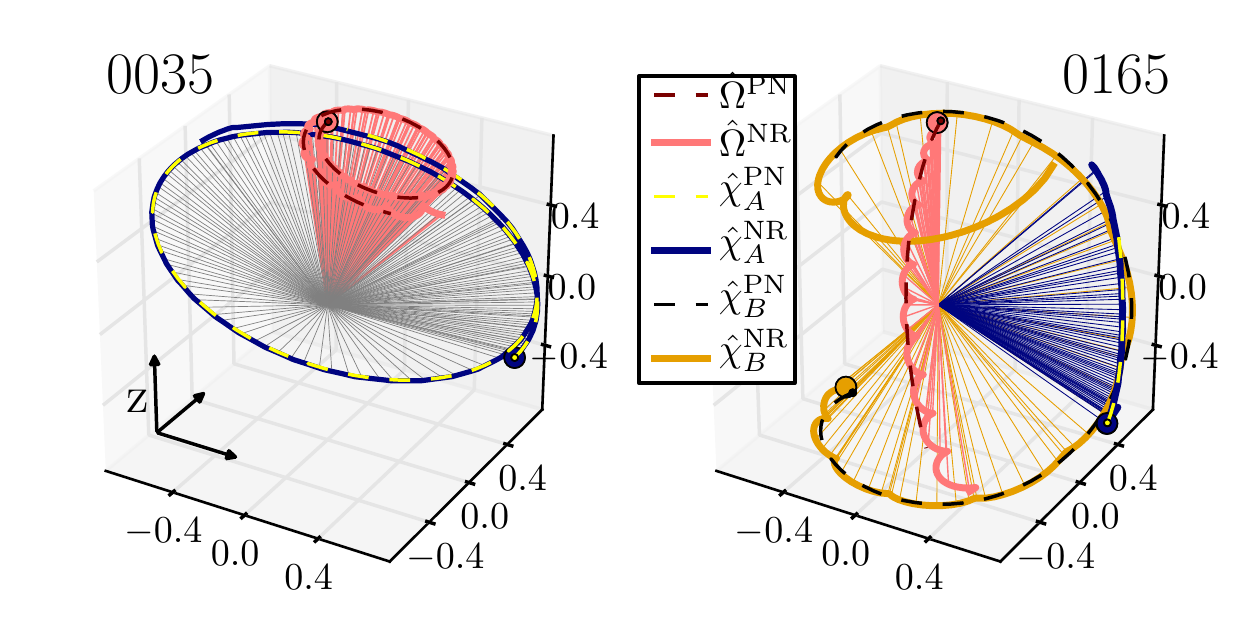}
\includegraphics[width=\columnwidth]{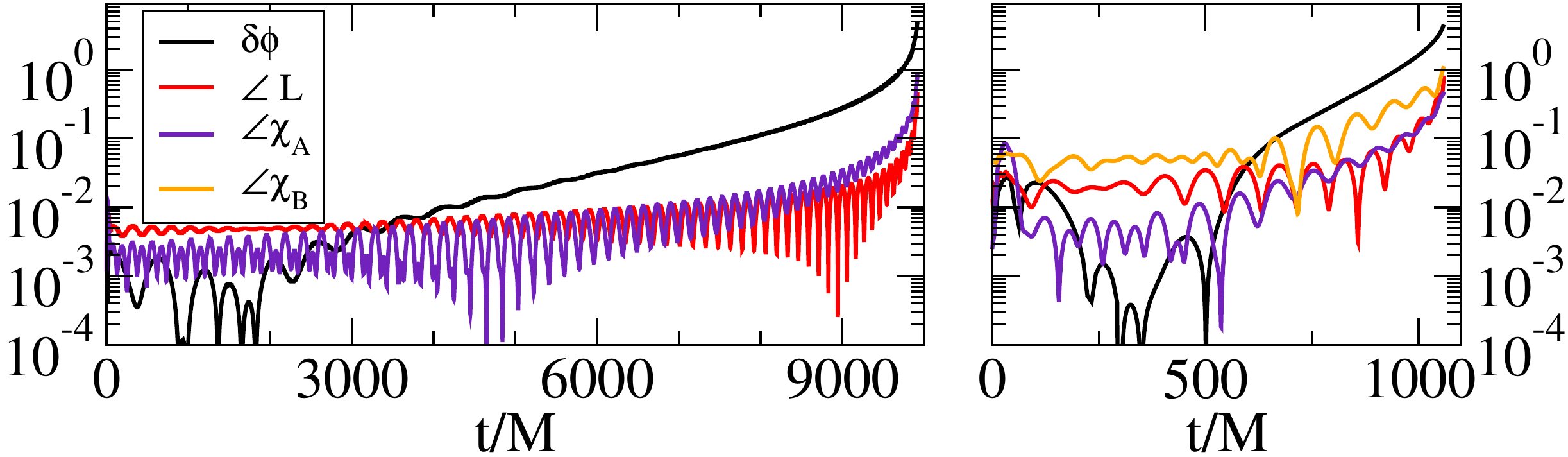}
\caption{\label{fig:trajectory-spin} Precessing simulations $0035$ and $0165$.
  {\bf Top:} Precession and nutation of the orbital angular momentum
  and the spin vectors of the BHs.  The circles denote the initial
  location of the vectors. {\bf Bottom:}
      Comparison of dynamical quantities
    with post-Newtonian theory: angle $\angle L$ between PN and NR
    Newtonian orbital angular momentum; angle $\angle \chi_{A,B}$
    between PN and NR spins ($\chi_B=0$ for 0035); 
    difference $\delta\phi_{\rm orb}$ in orbital phase.  
    All differences are in radians.
  }
  \end{figure}

\begin{figure}
\includegraphics[width=\columnwidth]{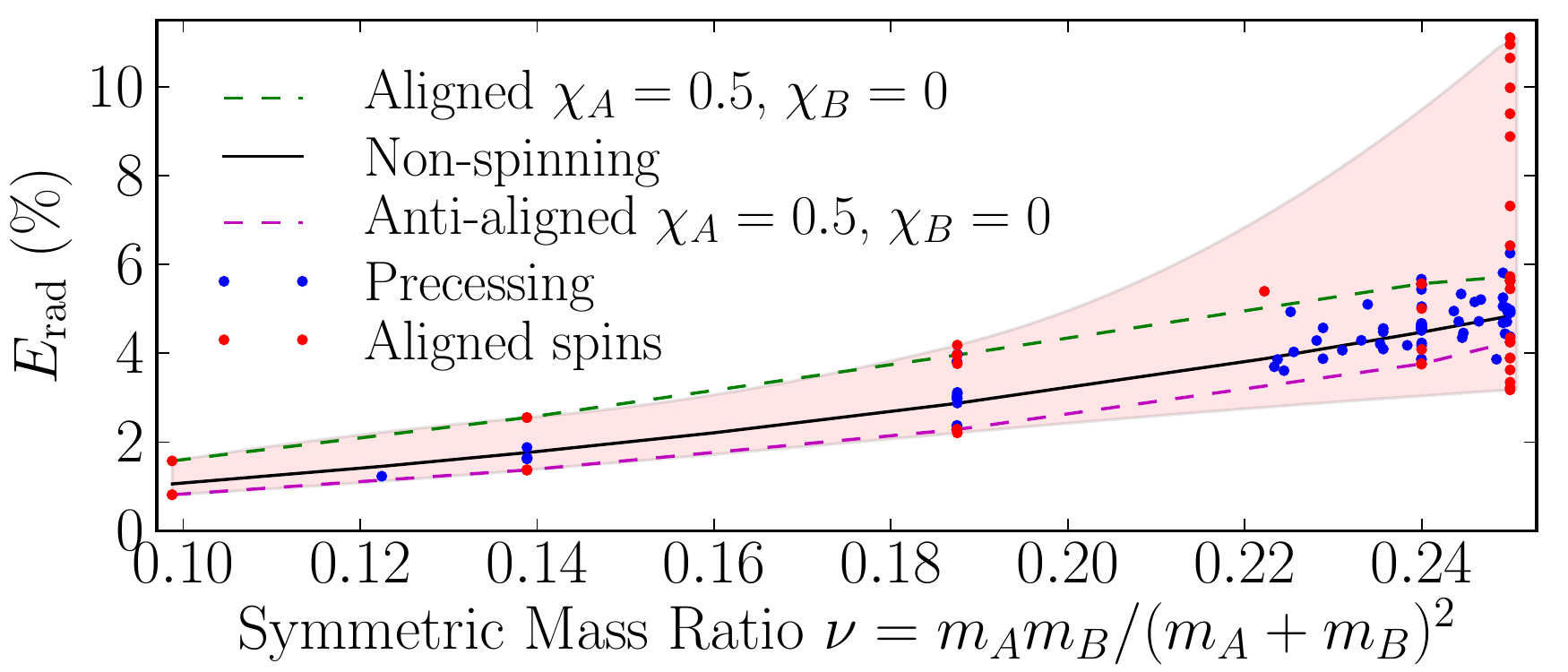}
\caption{The radiated energy (as measured by the fractional difference 
between the initial and final Christodoulou mass, where the initial mass 
is the sum of the holes' masses after the initial relaxation) for each 
simulation in the catalog.
\label{fig:radiatedEnergy}
}
\end{figure}

\section{Discussion}

This paper presents the most comprehensive catalog of high-quality BBH
simulations to date, 
 enabling studies
to help maximize the impact of GW detectors and to increase our understanding
of GW sources and dynamical, strongly curved spacetime:

\emph{Accuracy of PN precession equations:} PN
theory predicts how spin and orbital angular momenta 
precess in generic binaries 
(e.g.~\cite{2013arXiv1302.6723H}).  
%
%
The simulations here are long enough for detailed comparisons at
different points in parameter space.  As an
  example, Fig.~\ref{fig:trajectory-spin} highlights two precessing
  systems.  (i) Simulation 0035 
  (cf Fig.~\ref{fig:convergence}) 
  proceeds through about 1.5
  precession cycles: the normal to the instantaneous orbital plane
  traces out the red precession cone with opening angle $35^\circ$,
  whereas the spin of the more massive 
  BH traces out a
  precession cone with an opening angle of $144^\circ$.
Simulation 0165 has $q=6$ 
and spin magnitudes $\chi_A=0.9$, $\chi_B=0.3$ in generic directions.  
The orbital plane changes
by almost $90^\circ$, and the spin direction of the smaller 
BH traces out a spiral motion. 
The orbital angular 
momentum and spin vectors computed from the most recent available
PN approximant~\cite{2013arXiv1303.7412B} 
(dashed lines) agree well with numerical relativity (NR)
until the end of the inspiral.
The lower panel of Fig.~\ref{fig:trajectory-spin} 
quantifies this agreement. 
Details of this study will be discussed in~\cite{Inprep-Ossokine-et-al:2013}.  

\emph{Precessing inspiral-merger-ringdown waveform
  models} \cite{Sturani:2010yv,Pan:2013rra,Hannam:2013oca} can be tested and improved with the catalog, and new models can be developed. 
  Some of these models require knowledge of radiated energy; 
  we investigate this
  in Fig.~\ref{fig:radiatedEnergy}.  
  More energy is radiated at lower mass ratios, and
  the simulation with the most radiated energy ($11.1\%$, the highest
  to date for BBH inspirals) is new, 
  having the highest black-hole spins ($\chi = 0.98$)
  to date. This radiated energy is larger than the maximum 
  predicted by previous phenomenological fits 
  (compare \cite{Reisswig:2009vc,Barausse:2012qz}) but 
  agrees well with the recent fitting formula of 
  Ref.~\cite{Hemberger:2013hsa}, 
  which includes simulations with spins up to $\chi=0.97$.
  To our knowledge, no previous remnant mass
  calculations have been performed for spinning BBHs at mass ratios
  $q>4$.

  \emph{Periastron advance} can be studied in aligned spin binaries
  and generic binaries~\cite{LeTiec-Mroue:2011,Mroue2010,Hinderer:2013uwa}, 
  using simulations with different eccentricities.  
  This allows the first calculation of the
  gravitational self-force contribution to the periastron advance for
  spinning binaries~\cite{Tiec:2013twa}.

\emph{PN accuracy studies, extended to a larger region of
  parameter space and to an earlier stage of the inspiral:} Previous
studies consider only aligned
spins~\cite{Hannam:2010ec,Hannam2007c,Lovelace:2011nu} and at best either
15--20 orbits~\cite{MacDonald:2012mp,Boyle2007} or a few orbits at 
large separation~\cite{Lousto:2013oza}.  With the longer waveforms in
this catalog, these studies can extend to the earlier part of the
inspiral, where PN theory is expected to be more accurate 
and to include a larger region of parameter space.

\emph{Independent validation of existing analytical waveform models:}
Many waveform
models~\cite{Damour:2008te,Ajith2009,Pan:2009wj,Buonanno:2009qa,Santamaria:2010yb,Taracchini:2012}
are calibrated against numerical relativity simulations---but usually
only with a small number of short (typically $<10$ orbits) 
simulations.  The new simulations here enable tests of
these models at many different points in parameter space and
covering more cycles.

\emph{Detection sensitivity:} Following the approach taken in the
NINJA projects~\cite{Ajith:2012az,Aylott:2009ya,Aylott:2009tn}, our waveforms 
can be injected into GW detector noise to study the efficiency of GW 
data-analysis pipelines. Injections of precessing and/or eccentric waveforms
from this catalog can quantify the
impact of precession and eccentricity on the detection sensitivity of 
current searches using circular, aligned-spin templates.  The new
waveforms will also help assess the performance of
searches with precessing waveform templates.

\emph{Systematic errors in parameter estimation:}
Parameter estimation methods~\cite{Aasi:2013kqa} currently 
use inspiral-only PN waveforms.
Applying parameter estimation methods to 
the waveforms in this catalog 
will enable the systematic errors of this approach to be quantified.

While this catalog will enable pioneering studies,
major challenges remain for future work. First, for a
waveform to be most useful for data analysis, it must be connected to
a PN waveform from the early inspiral, forming a {\em hybrid
  waveform}~\cite{Ajith:2012az} that spans the entire frequency range
of a GW detector.  This is difficult for precessing configurations,
because of both the complexity of precessing PN waveforms and
ambiguities in connecting PN binary parameters with the numerical
binary parameters~\cite{Boyle:2013a}.

Second, 
most of the parameter space remains unexplored.  
Only 24
configurations have mass ratio $q>3$ (cf. Fig.~\ref{fig:listofruns}); of these, 
only 5 are precessing, and almost none have
a spinning smaller black hole.  Spinning BBH systems for
$5\lesssim q\lesssim 10$ are particularly interesting 
because they may serve as accurate proxies for BH-NS
binaries~\cite{Foucart:2013psa}. 
Furthermore, the catalog contains
only four simulations (the only four to date~\cite{Hemberger:2013hsa,Lovelace:2011nu})
with spins $\chi\ge 0.93$.

Finally, for some simulations, the numerical resolution is determined
by specifying the adaptive mesh refinement (AMR) tolerance. The
times at which AMR makes adjustments depend on this tolerance,
so comparing simulations with different AMR tolerances 
is not always as straightforward 
to interpret as Fig.~\ref{fig:convergence}.
This will be discussed within a more
comprehensive analysis of this catalog in~\cite{Mroue:2013InPrep}.
The catalog is available for download~\cite{SXSCatalog}.

\EarlyEnd
\begin{acknowledgments}

We thank Christian Ott and Kip Thorne
for helpful discussions.  This work was supported in part by 
NSERC of Canada, the Canada Chairs Program, 
and the Canadian Institute for Advanced Research;
the Sherman Fairchild Foundation; NSF
grants PHY-0969111 and PHY-1005426 at Cornell, NSF grants
PHY-1068881, PHY-1005655, and DMS-1065438 at Caltech, and 
NSF grant PHY-1307489 at Cal State Fullerton.  
Simulations used in this work
were computed with the \texttt{SpEC} code~\cite{SpECwebsite}.  Computations
were performed on the Zwicky cluster at Caltech, which is supported by
the Sherman Fairchild Foundation and by NSF award PHY-0960291; on the
NSF XSEDE network under grant TG-PHY990007N; 
on the Orca cluster supported by Cal State Fullerton; 
and on the GPC
supercomputer at the SciNet HPC Consortium~\cite{scinet}. SciNet is
funded by: the Canada Foundation for Innovation under the auspices of
Compute Canada; the Government of Ontario; Ontario Research
Fund--Research Excellence; and the University of Toronto. 
\end{acknowledgments}

\bibliography{../References/References}

\end{document}